\documentclass[prd,showpacs,twocolumn,
amsmath,amssymb,superscriptaddress,floatfix,nofootinbib]{revtex4-1}
\usepackage[utf8]{inputenc}
\usepackage[sort&compress]{natbib}
\usepackage[normalem]{ulem}
\usepackage{lipsum} 
\usepackage{bm}
\usepackage{times}
\usepackage{amssymb,amsbsy,amsmath,amsfonts}
\usepackage{graphicx}
\usepackage{float}
\usepackage{color}
\usepackage{morefloats}
\usepackage{rotating}
\usepackage{srcltx}
\usepackage{slashed}
\usepackage{subfigure}
\usepackage{multirow}
\usepackage{verbatim}
\usepackage{hyperref}
\usepackage{tabularx}
\usepackage{adjustbox}
\usepackage[dvipsnames]{xcolor}
\usepackage{nicefrac}

\usepackage{hyperref}
\hypersetup{
	colorlinks	=true,
	urlcolor	=blue,
	linkcolor	=blue,
	citecolor	=blue,
	pdftitle	={},
	pdfauthor	={Zhi-Wei Liu, Jun-Xu Lu, Ming-Zhu Liu, Li-Sheng Geng},
	pdfsubject	={$\Sigma_c\bar{D}^{(*)}$ correlation function}
	}

\begin{document}

\title{Implication of a negative effective range on the $D\bar{D}^*$ interaction and the nature of $X(3872)$
}

\author{Yi-Bo Shen}
\affiliation{School of Physics, Beihang University, Beijing 102206, China}

\author{Ming-Zhu Liu}
\affiliation{
Frontiers Science Center for Rare Isotopes, Lanzhou University,
Lanzhou 730000, China}
\affiliation{ School of Nuclear Science and Technology, Lanzhou University, Lanzhou 730000, China}

\author{Zhi-Wei Liu}
\email[Corresponding author: ]{liuzhw@buaa.edu.cn}
\affiliation{School of Physics, Beihang University, Beijing 102206, China}

\author{Li-Sheng Geng}
\email[Corresponding author: ]{lisheng.geng@buaa.edu.cn}
\affiliation{School of Physics, Beihang University, Beijing 102206, China}
\affiliation{Peng Huanwu Collaborative Center for Research and Education, Beihang University, Beijing 100191, China}
\affiliation{Beijing Key Laboratory of Advanced Nuclear Materials and Physics, Beihang University, Beijing 102206, China }
\affiliation{Southern Center for Nuclear-Science Theory (SCNT), Institute of Modern Physics, Chinese Academy of Sciences, Huizhou 516000, China}

\begin{abstract}

A recent analysis of the LHCb data \href{https://journals.aps.org/prd/abstract/10.1103/PhysRevD.105.L031503}{[Phys. Rev. D 105 (2022) L031503]} obtained a sizable negative effective range for the $X(3872)$. This has attracted intensive discussions on whether
$X(3872)$ can be deemed as a $D\bar{D}^*$ molecular state. This work explicitly demonstrates that the negative effective range of the $X(3872)$ does not contradict 
the molecular picture, adopting an effective field theory formulation of the $D\bar{D}^*$ interaction that can simultaneously reproduce the binding energy and effective range of the $X(3872)$.
We elaborate on the implications of the large negative effective range of $X(3872)$ and the small binding energy on the underlying $D\bar{D}^*$ interaction. Such results are relevant for 
a better understanding of hadronic molecules and their binding mechanism. 
\end{abstract}


\maketitle

\section{Introduction}

The number of so-called exotic hadrons, beyond the configurations of mesons made of a pair of quark and antiquark and baryons made of three quarks in the conventional quark model~\cite{Gell-Mann:1964ewy, Zweig:1964ruk}, has been increasing~\cite{Briceno:2015rlt,Richard:2016eis,Yuan:2018inv,Belle-II:2018jsg,BESIII:2020nme,Gershon:2022xnn,Liu:2023hhl}, which not only enrich hadron spectroscopy but also provide valuable opportunities to understand the non-perturbative strong interactions better.  Despite the intensive experimental and theoretical studies, the nature of these exotic states remains controversial~\cite{Chen:2016qju,Lebed:2016hpi,Oset:2016lyh,Esposito:2016noz,Dong:2017gaw,Guo:2017jvc,Olsen:2017bmm,Ali:2017jda,Karliner:2017qhf,Guo:2019twa,Brambilla:2019esw,Liu:2019zoy,Meng:2022ozq,Liu:2024uxn}. Because most of them are located close to the mass thresholds of pairs of conventional hadrons, the hadronic molecular picture has become very popular~\cite{Guo:2017jvc,Liu:2024uxn}. Recent studies have shown that in addition to the hadronic molecular components, other configurations, such as compact multiquark components, also contribute to the experimentally discovered states, leading to complicated structures for these states~\cite{BESIII:2023hml,LHCb:2020xds}.               

 To  estimate the relative importance of  the hadronic molecular component in a physical state,  one often turns to the Weinberg compositeness criterion, which defines the probability of finding an elementary component in the physical state corresponding to the field renormalization constant $Z$~\cite{Weinberg:1962hj}, 
\begin{eqnarray}
 Z= 1- \int {\rm d} \alpha |\langle \alpha | d \rangle |^2,~~~ Z=\sum_{n}|\langle n | d \rangle |^2, 
\end{eqnarray}
where $| \alpha \rangle$ and  $| n \rangle$ represent the eigenstates of the continuum and discrete elementary particle states in the free Hamiltonian $H_0$, and $| d \rangle$ represents the physical state in the total Hamiltonian $H$ with the normalization of $\sum_n | n \rangle \langle n| + \int {\rm d} \alpha | \alpha \rangle \langle \alpha | =1$ and $\langle d| d \rangle =1$. $Z=0$ implies that the physical state is a pure hadronic molecule, and $0<Z<1$ indicates the existence of an elementary component inside the physical state.

One can relate $Z$ to the scattering amplitude at low energies in a model-independent way, which can be expressed by the effective range  expansion (ERE)
\begin{eqnarray}
 f(k)=\frac{1}{k\cot \delta - ik}\approx \frac{1}{-\frac{1}{a}+\frac{1}{2}r_0 k^2 - ik},  \label{ERE Formula} 
\end{eqnarray}
where $\delta$ is the phase shift, $a$ is the scattering length, $r_0$ is the effective range, $k=\sqrt{2\mu E}$, and $\mu$ is the reduced mass. The scattering length and effective range can be expressed in terms of $Z$ as~\cite{Baru:2021ldu}
\begin{eqnarray}
 a= \frac{2(1-Z)}{2-Z}\frac{1}{\gamma} +\mathcal{O}(\frac{1}{\beta}), ~~~~~  r_0=-\frac{Z}{1-Z}\frac{1}{\gamma}+\mathcal{O}(\frac{1}{\beta}), \label{eq:weinberg}    
\end{eqnarray}
where $\gamma=\sqrt{2\mu B}$ is the binding momentum,  and $1/\beta$ estimates the range corrections. From the relation above, following Ref.~\cite{Matuschek:2020gqe} and ignoring higher order corrections, one can obtain the compositeness $X=1-Z$, which can be seen as the probability of finding the hadronic molecular component in the normalized wave function of the bound state
\begin{equation}
    X=\sqrt{\frac{a}{a-2r_{0}}}.
    \label{X}
\end{equation}
Since   $\beta$ denotes the next momentum scale not treated explicitly in the ERE, it is normally regarded as the mass of the lightest exchanged particle.     

Applying this rule to the deuteron, a bound state composed of one proton and one neutron, one obtains a compositeness $X=1.68$
with the following experimental values for the scattering length, effective range, and binding energy: $
 a=5.419(7)$ fm,  $r_0=1.766(8)$ fm,  and $B=-2.224575(9)$ MeV. The result is unacceptable because $X$ should be smaller than one by definition. As stressed in Ref.~\cite{Song:2022yvz}, the alternative is even worse. As a result,  Weinberg concluded, ``The true token
that the deuteron is composite is that $r_0$ is small and positive rather than large and negative~\cite{Weinberg:1965zz}."

Many studies of exotic hadrons employed the Weinberg compositeness criterion to classify a particular hadron as either a molecular or nonmolecular state~\cite{Albaladejo:2022sux,Baru:2021ldu,Kinugawa:2023fbf,Montesinos:2024uhq,Esposito:2021vhu,Mikhasenko:2022rrl,Kinugawa:2022fzn}. The $X(3872)$ is the most studied among the many exotic hadrons discovered. However, whether it is a compact tetraquark state or a loosely bound molecule remains unsettled~\cite{Maiani:2007vr,Esposito:2021vhu,BESIII:2022qzr,Kinugawa:2022fzn,Wu:2023rrp,Grinstein:2024rcu}. By fitting the LHCb data~\cite{LHCb:2020xds}, Ref.~\cite{Esposito:2021vhu} obtained an effect range for the $X(3872)$, $r_{0}=-5.34$ fm~\footnote{After subtracting the contribution from the second
channel, a reanalysis of the same data yielded a value of about $-3.78$ fm~\cite{Baru:2021ldu}.}, and an inverse scattering length $\kappa_{0}=6.92$ MeV. According to the Weinberg relation of Eq.~(\ref{eq:weinberg}) and the Landau relation:
\begin{equation}
  Z=\frac{-r_{0} \kappa}{1-r_{0} \kappa},
\end{equation}
they obtained $Z=0.14$ and concluded that the $X(3872)$ cannot be a pure shallow molecule, which means $Z=0$ and the effective range must be strictly positive, as happens for the deuteron. We note, however, that most studies would classify such a state with $Z=0.14$ (i.e., $X=0.86)$ as a hadronic molecule. They further stated that only if the $X(3872)$ is an elementary object, the effective range is negative, and its magnitude is much larger than that of the inverse pion mass.

It should be noted that when applied to hadronic molecules, the Weinberg composition criterion should be corrected since the dynamics of hadronic molecules are more complicated. Recently, the impact of the range of the effective potentials to the effective range has been extensively discussed~\cite{Baru:2021ldu,Kinugawa:2021ykv,Albaladejo:2022sux,Song:2022yvz}. As argued in Ref.~\cite{Baru:2021ldu}, because the one-pion exchange is responsible for the nucleon-nucleon scattering, the corresponding value of $1/\beta$ is  $1.4$~fm. In contrast, the value for a pair of charmed mesons is reduced to  $1$~fm because of the suppression of the one-pion exchange contribution. Moreover, the authors stressed that coupled-channel effects, isospin violation effects, and the widths of constituents also modify the effective range of $X(3872)$. Considering the contributions from the next-to-leading order Weinberg relation, Ref.~\cite{Albaladejo:2022sux} confirmed that the deuteron is a composite particle and concluded that higher-order corrections of the effective range expansion are required for deeply bound states.

In this work, we want to understand what the effective range can tell about the underlying hadron-hadron interactions, e.g., the $D\bar{D}^*$ interaction for the $X(3872)$. In particular, we study two types of $D\bar{D}^*$ interactions and check whether they can reproduce the binding energy and negative effective range of the $X(3872)$ and simultaneously yield a compositeness $X$ greater than 0.5, following the criterion for a hadronic molecule adopted by most studies. We note that unlike most previous studies trying to quantify to what extent a specific state can be deemed as a molecular state or nonmolecular state~\cite{Albaladejo:2022sux,Song:2022yvz,Baru:2021ldu,Kinugawa:2023fbf,Montesinos:2024uhq,Esposito:2021vhu,Mikhasenko:2022rrl,Kinugawa:2022fzn,BESIII:2022qzr,Kinugawa:2022fzn,Wu:2023rrp,Grinstein:2024rcu},  we try to determine what the effective range and the binding energy can tell about the underlying hadron-hadron interaction. In this sense, our present work is similar in spirit to Ref.~\cite{Song:2022yvz} but with different focuses. 

This work is organized as follows. In Sect. II, we explain how to
fix the potential parameters, solve the Lippmann-Schwinger equation in momentum space, and calculate the binding energy, scattering length, effective range, and compositeness of $X(3872)$. Results and discussions are given in Sect.III, followed by a summary in the last section.

\begin{figure}
\includegraphics[width=0.45\textwidth]{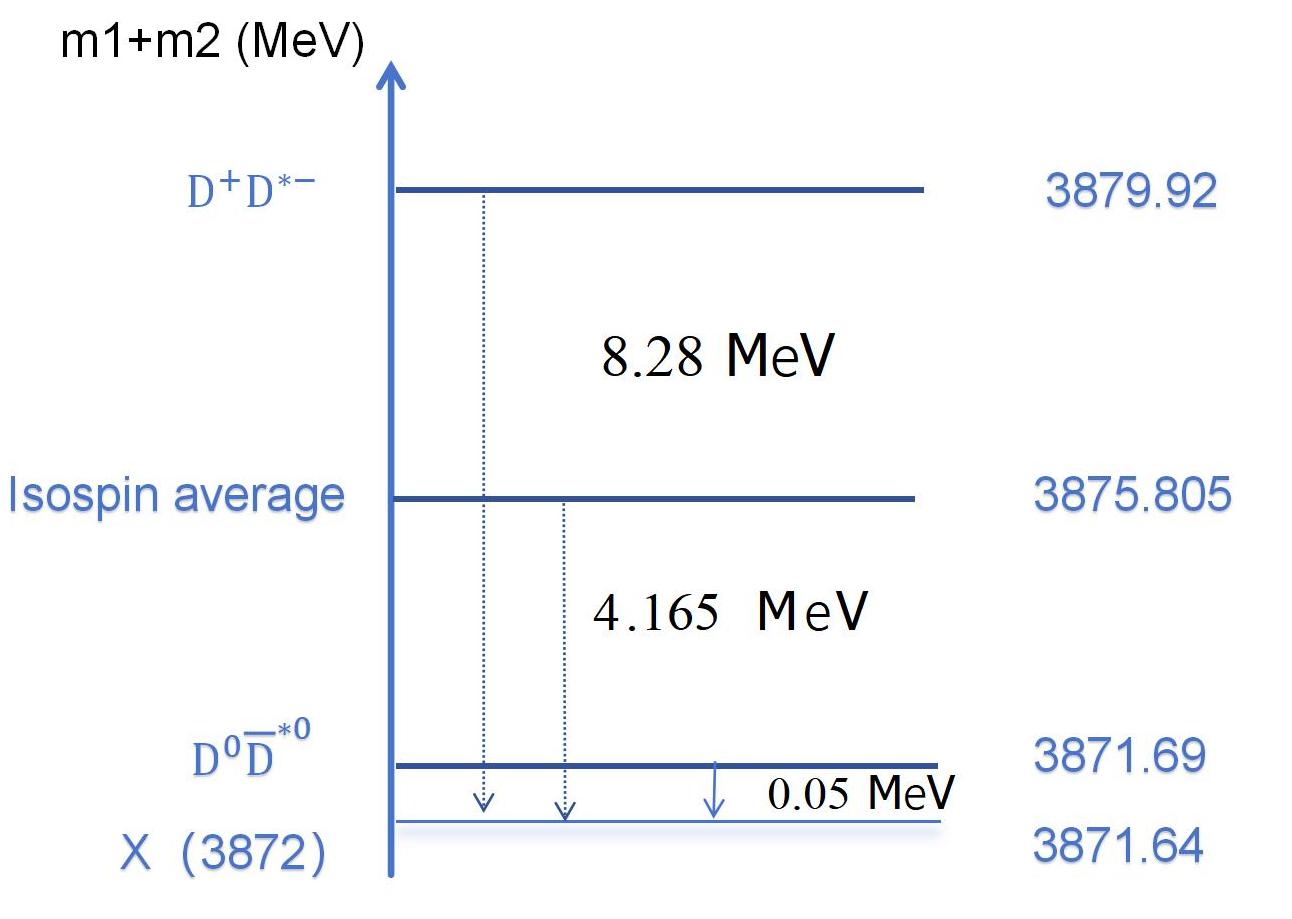} 
\caption{Schematic plot of the thresholds of $D^0\bar{D}^{*0}$, $D^+D^{*-}$, their isospin average, and the location of the $X(3872)$.  \label{schemactic}
}
\end{figure}

\section{Theoretical Formalism}
To better understand the relation between the effective range and the underlying $D\bar{D}^*$ interaction, we work in momentum space, which allows one to parameterize the $D\bar{D}^*$ interaction in a model-independent way.  
We study first the single-channel $D\bar{D}^*$ interaction to simplify the discussion without losing generality. In this case, we use the masses of their neutral partners for the masses of $D$ and $\bar{D}^*$. In the particle basis, the binding energy of the $X(3872)$ is only -0.018 MeV with respect to the threshold of $D^0\bar{D}^{*0}$~\cite{ParticleDataGroup:2020ssz}. Working with the isospin averaged masses, the $X(3872)$ binding energy would be about $-4$ MeV~\cite{Wu:2023rrp}. We will study how such a large binding energy affects our understanding. We then study the coupled-channel case to check the robustness of our conclusion. For the sake of reference,  we show in Fig.~\ref{schemactic} the mass thresholds of the neutral channels, charged channels, and isospin channels of $D\bar{D}^*$ relative to the mass of $X(3872)$.

In momentum space, close to the threshold,  the $D\bar{D}^*$ interaction can be parameterized as
\begin{equation}
    V=\alpha+\beta k^{2},
\end{equation}
where $\alpha$ and $\beta$ are low-energy constants (LECs) to be determined. With this potential, one can solve the following Lippmann-Schwinger
equation to search for poles, i.e., 
\begin{equation}
    T(s)=\frac{V(s)}{1-V(s)\cdot G(s)},
\label{T}
\end{equation}
where $G(s)$ is the two-point one-loop function:
\begin{equation}
\begin{split}
    G(s)&=\int_{0}^{q_\mathrm{max}}\frac{{\rm d}q~q^{2}}{4\pi^{2}}\frac{\omega(m_{1},q)+\omega(m_{2},q)}{\omega(m_{1},q)\cdot\omega(m_{2},q)}\\
    &\times\frac{1}{s-[\omega(m_{1},q)+\omega(m_{2},q)]^{2}+i\epsilon},
\end{split}
\label{loop function}
\end{equation}
and $\omega(m_{1,2},q)=\sqrt{m_{1,2}^{2}+q^{2}}$ is the energy of $D$ and $\bar{D}^*$, $\sqrt{s}$ is the center-of-mass energy of the $D\bar{D}^*$ system, $q_\mathrm{max}$ is a sharp cutoff momentum to regulate the logarithmically divergent loop function~\cite{Ali:2019npk}, and $q$ is the center-of-mass momentum of $D$ and $\bar{D}^*$.  From the $T$ matrix at threshold, one can obtain the scattering length $a$ and effective range $r_0$ as follows~\cite{Ikeno:2023ojl}:
\begin{equation}
    -\frac{1}{a}=-\left.8 \pi \sqrt{s} T^{-1}\right|_{s=s_{\mathrm{th}}},
\end{equation}
\begin{equation}
\begin{aligned}
r_{0} & =\frac{\partial}{\partial k^{2}} 2\left(-8 \pi \sqrt{s} T^{-1}+i k\right) \\
& =\left.\frac{\sqrt{s}}{\mu} \frac{\partial}{\partial s} 2\left(-8 \pi \sqrt{s} T^{-1}+i k\right)\right|_{s=s_{\text {th }}},
\end{aligned}
\end{equation}
where $s_\mathrm{th}=(m_1+m_2)^2$.
With $a$ and $r_{0}$ obtained, one can calculate the compositeness $X$ following Ref.~\cite{Song:2022yvz}~\footnote{We do not use the Weinberg relation due to its shortcomings mentioned in the introduction.}:
\begin{equation}
    X = 1 - \left.\frac{1}{
         \frac{\partial V^{-1}}{\partial s}  
        - \frac{\partial G}{\partial s} }
         \frac{\partial V^{-1}}{\partial s}\right|_{s_0}, 
    \label{X2}
\end{equation}
where $s_{0}$ is the pole position.

Next, we consider the coupled-channels of $D^0\bar{D}^{*0}$ and $D^+D^{*-}$. The $X(3872)$ wave function as an isospin zero state is given by
\begin{equation}
    \left|D\bar{D}^{*} , I=0\right\rangle=\frac{1}{\sqrt{2}}\left(D^0 \bar{D}^{*0}+D^+ D^{*-}\right).
\end{equation}
If we strictly follow the effective field theory approach, we will have too much freedom in choosing the coupled-channel potential, which reads
\begin{equation}
    \tilde{V}_{R}=\left(\begin{array}{ll}
V_D& V_C \\
V_C & V_D
\end{array}\right).
\end{equation}
Namely, we will need four LECs, two for $V_C$ and two for $V_D$. As a result, without loss of generality, we turn to phenomenological models for guidance to avoid introducing too many LECs. The hidden-gauge theory~\cite{Song:2023pdq} tells that $V_C=V_D$. Therefore, we can assign $V_C=V_D=\frac{1}{2}(\alpha+\beta k^2)$.

It is straightforward to obtain the $T$ matrix in this coupled-channel case with the above potential and the following loop function in matrix form
\begin{equation}
    G=\left(\begin{array}{cc}
G_{D^{0} \bar{D}^{* 0}} & 0 \\
0 & G_{D^+ D^{*-}}
\end{array}\right).
\end{equation}
In the coupled-channel case, we will have two scattering lengths, two effective ranges, and two compositenesses, one for each channel. They read explicitly as 



 
\begin{align}
    -\frac{1}{a_{1}} &= \left.(-8 \pi \sqrt{s})T_{11}^{-1}\right|_{s_{\mathrm{th} 1}},
    \label{a1}
\end{align}

\begin{align}
    r_{0,1}= & \left.2 \frac{\sqrt{s}}{\mu_{1}} \frac{\partial}{\partial s} ( - 8 \pi \sqrt { s } ) T_{11}^{-1}\right|_{\mathrm{s}_{\mathrm{th} 1}},
    \label{r1}
\end{align}

\begin{align}
    -\frac{1}{a_{2}} &= \left.(-8 \pi \sqrt{s})T_{22}^{-1}\right|_{s_{\mathrm{th} 2}},
    \label{a2}
\end{align}

\begin{align}
    r_{0,2}= & \left.2 \frac{\sqrt{s}}{\mu_{2}} \frac{\partial}{\partial s} ( - 8 \pi \sqrt { s } ) T_{22}^{-1}\right|_{\mathrm{s}_{\mathrm{th} 2}},
    \label{r2}
\end{align}

\begin{equation}
    P_{1}=\left.-\lim _{s \rightarrow s_{0}}\left(s-s_{0}\right) T_{11}\frac{\partial G_{1}}{\partial s}\right|_{s_{0}},
\end{equation}

\begin{equation}
    P_{2}=\left.-\lim _{s \rightarrow s_{0}}\left(s-s_{0}\right) T_{22}\frac{\partial G_{2}}{\partial s}\right|_{s_{0}},
\end{equation}
where $P_{1}$,$P_{2}$ represent the probability to find the hadronic molecular component in the channel 1 and channel 2.

\begin{figure*}[htpb]
  \centering
  \includegraphics[width=\textwidth]{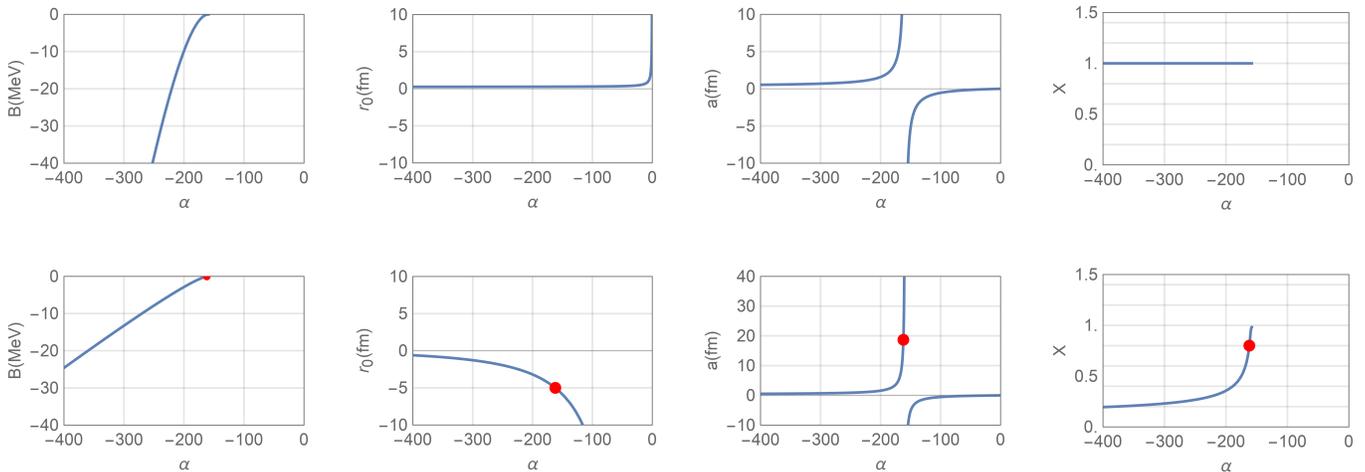}
  \caption{Variations of the binding energy, scattering length, effective range, and compositeness for two types of potentials in momentum space $V=\alpha$ (upper), 
 and $V=\alpha+\beta k^2$ with fixed $\beta_0$ (lower). The red points represent the solution for the $X(3872)$, $a=18.7$ fm and $X=0.8$, with $\alpha_0=-162$ and $\beta_0=-3616$ GeV$^{-2}$.\label{fig;single}}
\end{figure*}

\begin{figure*}[htpb]
  \centering
  \includegraphics[width=\textwidth]{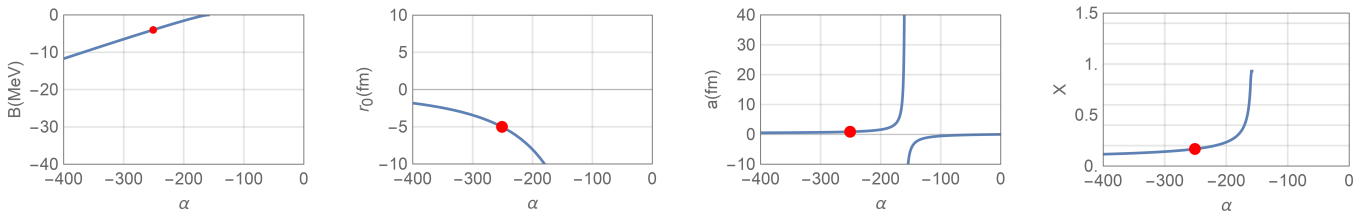}
  \caption{The same as Fig.~\ref{fig;single}, but for a binding energy of $-4$ MeV. The solution for the $X(3872)$ yields 
$a=0.88$  and $X=0.17$ with $\alpha_0=-251$ and $\beta_0=-8650$ GeV$^{-2}$.
 \label{fig;single2}}
\end{figure*}

\section{Results and discussions}
\begin{table}[ht]
    \centering
        \caption{Value of physical quantities relevant to the present work, where $B$ is the binding energy of the $X(3872)$ and $r_0$ is its effective range~\cite{Esposito:2021vhu}. The binding energy and masses are in units of MeV and the effective range is in units of fm.}
    \label{tab:parameters}
    \resizebox{\linewidth}{!}{
    \begin{tabular}{c|c|c|c|c|c}
        \hline\hline
        B & $D^{0}$ & $\overline{D}^{*0}$ & $D^{+}$& $D^{*-}$ & $r_{0}$  \\
        \hline
        $-0.05$& 1864.84& 2006.85 &  1869.66 & 2010.26& $-5.34 < r_{0} < -3$ \\
        \hline\hline
    \end{tabular}
    }
\end{table}
In this section, we study the binding energy, scattering length, effective range, and compositeness of $X(3872)$ and 
check whether it can be understood as a $D\bar{D}^*$ molecule with a negative effective range. 

In addition to the LECs $\alpha$ and $\beta$, the momentum cutoff $q_\mathrm{max}$ also needs to be determined, for which we choose a value of 1 GeV.\footnote{We have checked using a cutoff of 0.5 GeV barely affects the results and our conclusion.}  The relevant $D/\bar{D}^*$ masses, binding energy, and effective range are given in Table~\ref{tab:parameters}.

\begin{figure*}[htpb]
  \centering
  \includegraphics[width=\textwidth]{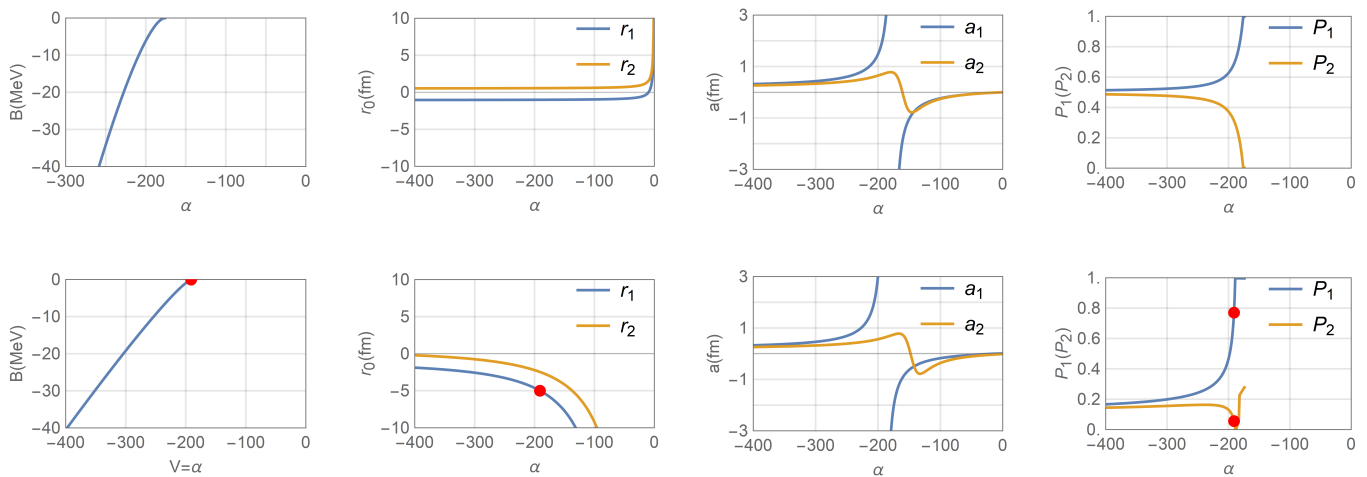}
  \caption{The same as Fig.~\ref{fig;single}, but for the coupled-channel case with $V_{12}=V_{21}=V_{11}=V_{22}$. The solution for the $X(3872)$ yields $a_1=15.7$ fm, $a_2=0.62$ fm, $p_1=0.77$, and $p_2=0.056$ with $\alpha_0=-191$ and $\beta_0=-1641$ GeV$^{-2}$.}
  \label{fig:couple1}
\end{figure*}


\subsection{Single-channel case}
We first focus on the single-channel case. The results for the case of $V=\alpha$ are shown in the upper panel of Fig.~\ref{fig;single} as
a function of $\alpha$.
One can see that the effective range $r_0$ is always positive, and the corresponding $X$ is equal to 1 by definition, as can be easily seen from Eq.~(\ref{X2}). In such a case, the potential energy $V=\alpha$ corresponds to a $\delta$ potential in coordinate space. Therefore, the interaction range is small. 

In order to obtain a negative effective range $r_{0}$, we turn to the next-to-leading order potential, i.e., $V=\alpha+\beta k^{2}$.
With such a potential, one can find solutions for the $X(3872)$, i.e., the potential can simultaneously yield $B$ and $r_{0}$ consistent with the experimental data, which determine the LECs $\alpha_0$ and $\beta_0$. According to Ref.~\cite{Song:2022yvz}, we can calculate the compositeness $X$ with Eq.~\eqref{X2}. Taking the binding energy $-0.05$ MeV, the effective range $r_{0}$ = $-5$ fm,  we obtain $X=0.80$; taking the effective range $r_{0}$ = $-3$ fm, we obtain $X=0.86$. It is clear that in both cases, the $X(3872)$ can be viewed as a $D\bar{D}^*$ bound state.  With $\beta$ fixed at $\beta_0$, we can vary $\alpha$ and study
the variation of the binding energy, scattering length, effective range, and compositeness with $\alpha$. The results are shown in the bottom panel of Fig.~\ref{fig;single}. As the binding energy (absolute value) becomes larger, the effective range moves closer to zero, and the scattering length and the compositeness decrease. 

It is interesting to note that if we took the binding energy of the $X(3872)$ with respect to the isospin averaged mass threshold, i.e., about $-4$ MeV, we would have obtained the results shown in Fig.~\ref{fig;single2} (also using isospin averaged masses for $D$ and $\bar{D}^*$). The scattering length is less than 1 fm in such a case, while the compositeness $X$ becomes about 0.2. The compositeness is inversely related to the distance to the threshold, as more thoroughly studied in several studies~\cite{Song:2023pdq,Dai:2023kwv}.

\subsection{Coupled-channel case}
Next, we consider the coupled-channel case. Note that we follow the implication of the local hidden gauge approach. That is, the off-diagonal elements are the same as the diagonal elements. The results are shown in Fig.~\ref{fig:couple1}. Compared to the single-channel case, this figure is a bit complicated because there are two channels. As a result, there are two scattering lengths, two effective ranges, and two compositenesses. First, we note that even in the coupled-channel scenario, one cannot simultaneously reproduce the binding energy and negative effective range of the $X(3872)$ with a constant potential $V=\alpha$. As a result, as concluded from the single-channel scenario, a momentum-dependent potential is needed. 
Once the momentum dependence is considered, as seen from the bottom panel of Fig.~\ref{fig:couple1}, one can simultaneously reproduce the binding energy and effective range $r_1$ of the $X(3872)$. The corresponding scattering length $a_1$ is 16 fm, not far from the value of 19 fm in the single-channel scenario. The compositeness of channel 1 is 0.77,the state dominated by the $D^{0}\bar{D}^{*0}$, which is also consistent with the single-channel scenario. Namely, the $X(3872)$ can be largely viewed as a $D\bar{D}^*$ molecule. The variations of the effective range, scattering length, and compositeness are consistent with the single-channel results.

\subsection{Comparison with the $Z_c(3900)$}
It is interesting to compare the $X(3872)$ with the $Z_c(3900)$. They have long been believed to be isospin partners. In this work, we demonstrated explicitly that one needs a $D\bar{D}^*$ interaction of the form $\alpha+\beta k^{2}$ to reproduce the binding energy and effective range of the $X(3872)$. While to generate the $Z_c(3900)$ as a resonance above the $D\bar{D}^*$ threshold, the same form of the potential is needed~\cite{Wang:2020dko,Du:2022jjv,Liu:2024nac}. In addition, we note that in the one-boson exchange model, the $D\bar{D}^*$ interactions in the isospin zero and one channels are both induced by the exchange of $\sigma$, $\omega$ and $\rho$ mesons. However, the strength is larger for the isospin zero channel than for the isospin one channel~\cite{Sun:2011uh,He:2015mja,Liu:2019stu}. From this viewpoint, the results of the present study make sense.

Furthermore, we note that the so-called contact range effective field theories have been widely employed to connect the $X(3872)$ with various other systems~\cite{Liu:2020tqy,Dong:2021bvy,Chen:2021cfl,Peng:2021hkr}. How the momentum dependence of the potential affects these studies needs to be scrutinized.  

\section{Summary}
In this work, motivated by the debate on the nature of the $X(3872)$ from the perspective of its effective range, we studied whether
the $X(3872)$ can be dynamically generated as a $D\bar{D}^*$ molecular state and with the compositeness $X$ greater than 0.5. We adopted a model-independent parametrization of the $D\bar{D}^*$ potential. We showed
that the leading-order potential cannot accommodate a negative effective range, while the next-to-leading-order potential can. The resulting compositeness of the $X(3872)$ ranges from  0.8 (in the single-channel case) to about 0.77 in the coupled-channel case, indicating that it can indeed be viewed as a $D\bar{D}^*$ molecular state.

In addition to showing that a negative effective range is not an indicator of a non-molecular state, the present study provided invaluable clues on the $D\bar{D}^*$ interaction. The current experimental data show that the interaction needs to be the form of $\alpha+\beta k^2$ near the $D\bar{D}^*$ threshold. The exchange of light mesons can induce such a form in the single-channel case. This is easily achievable because the one-pion exchange for the $D\bar{D}^*$ interaction is allowed. Such an implication for various studies needs to be further scrutinized. 

Ref.~\cite{Baru:2021ldu} showed that coupled-channel hadron-hadron dynamics can naturally generate a large negative effective range. This is complementary to and consistent with the present study because even a constant contact interaction in a coupled-channel case, when reduced to a single channel, necessarily introduces energy (or momentum) dependence (see, e.g., Ref.~\cite{Song:2023pdq}), which is needed to account for the large and negative effective range.

\emph{Acknowledgments.} We thank Dr. Jing Song for early participation in this project and Prof. Eulogio Oset for the useful comments on the first draft of this manuscript. This work is partly supported by the National Key R\&D Program of China under Grant No. 2023YFA1606700 and the National Science Foundation of China under Grant No. 12435007. Zhi-Wei Liu acknowledges support from the National Natural Science Foundation of China under Grant No.12405133, No.12347180, China Postdoctoral Science Foundation under Grant No.2023M740189, and the Postdoctoral Fellowship Program of CPSF under Grant No.GZC20233381. Ming-Zhu Liu
acknowledges support from the National Natural Science Foundation of China under Grant No.12105007.

\bibliography{x3872}

\end{document}